\documentclass[12pt,preprint]{aastex}

\shorttitle{Mechanism of Outflows in Accretion System}
\shortauthors{Gu}

\begin{document}

\title{Mechanism of Outflows in Accretion System: \\
Advective Cooling Cannot Balance Viscous Heating?}

\author{Wei-Min Gu}

\affil{Department of Astronomy and Institute of Theoretical Physics
and Astrophysics, \\
Xiamen University, Xiamen, Fujian 361005, China}

\email{guwm@xmu.edu.cn}

\begin{abstract}
Based on no-outflow assumption, we investigate steady state,
axisymmetric, optically thin accretion flows in spherical coordinates.
By comparing the vertically integrated advective cooling rate
with the viscous heating rate, we find that the former is generally
less than 30\% of the latter, which indicates that the advective
cooling itself cannot balance the viscous heating.
As a consequence, for radiatively inefficient flows with
low accretion rates such as $\dot M \la 10^{-3} \dot M_{\rm Edd}$,
where $\dot M_{\rm Edd}$ is the Eddington accretion rate,
the viscous heating rate will be larger than the sum
of the advective cooling rate and the radiative cooling one.
Thus, no thermal equilibrium can be established
under the no-outflow assumption.
We therefore argue that in such case outflows ought to occur and
take away more than 70\% of the thermal energy generated by
viscous dissipation.
Similarly, for optically thick flows with extremely large
accretion rates such as $\dot M \ga 10 \dot M_{\rm Edd}$,
outflows should also occur due to the limited advection and
the low efficiency of radiative cooling.
Our results may help to understand the mechanism of
outflows found in observations and numerical simulations.
\end{abstract}


\section{Introduction}

Outflows may play an essential role in accretion system according to recent
observations and simulations. Based on Chandra observations
of Fe K$\alpha$ line, \citet{Wang13} found that outflows are significant
in our own Galaxy's supermassive black hole accretion system, and only
less than 1\% of the original gas can be accreted by the hole.
Apart from observations, some simulations also
showed that outflows exist in both optically thick and thin accretion systems.
For the optically thick, super-Eddington accretion case, \citet{Ohsuga05}
performed two-dimensional radiation-hydrodynamic simulations and found that
strong outflows exist and the inflow accretion rate is roughly proportional
to the radius. 
Such a scenario was confirmed by \citet{Ohsuga11} with two-dimensional
radiation-magnetohydrodynamic simulations, which also found that
outflows exist in optically thin, radiatively inefficient flows.
In addition, for the optically thin case, \citet{Yuan12} showed that
strong outflows exist and the inflow accretion rate can be described
by $\dot M \propto r^s$, where $s$ is in the range [0.4,~0.7].

Many mechanisms have been proposed to power outflows. For the optically
thick system, a possible mechanism is related to the radiation pressure.
With high accretion rates, the radiation force is so strong that the
vertical component of gravitational force cannot balance it.
Consequently, the excess radiation pressure will force a part of the
accreted gas into outflows. On the other hand, for the optically thin system,
the well-known advection-dominated accretion flow \citep[ADAF,][]{NY94}
may possess a positive Bernoulli parameter owing to its high internal energy,
which may account for outflows. Another mechanism is the BP process
\citep{BP82} with large-scale magnetic fields, which can work for both
optically thick and thin flows.

In the present work, we will study the mechanism for powering outflows
in a different way.
Our main concern is why outflows have to occur in many accretion systems.
During the past four decades there are three well-known and widely applied
accretion models \citep[for a review, see][]{Kato08},
namely the standard thin disk \citep{SS73}, the slim disk
\citep[][or the optically thick advection-dominated accretion flow]{Abram88},
and the optically thin advection-dominated accretion flow (ADAF).
The standard thin disk is radiative cooling dominated and it is believed
that outflows are quite weak in such a model.
On the contrary, the other two models are advective cooling dominated
and outflows are likely to be significant. Such a phenomenon
hints that there may exist some relationship between the outflows and
the strength of energy advection.

As argued in \citet{Gu07}, the classic slim disk model did not predict
outflows even for extremely high accretion rates since
a Taylor expansion of the vertical component of gravitational force
was adopted instead of the explicit one. Such an approximate force
will be greatly magnified for geometrically not thin disks and therefore
be able to balance the radiation force thus suppress outflows.
Once the explicit force is adopted, as shown in Figure~5 of \citet{Gu07},
no thermal equilibrium solution can exist for high accretion rates,
which implies that outflows are inevitable.
The physical reason for no solution is that the advective cooling
is not strong enough to balance the viscous heating.
Such a work is, however, limited to optically thick flows. Moreover,
some basic assumptions in this work, such as unified radial and azimuthal
velocity at a certain cylindrical radius $R$,
may not be appropriate particularly for geometrically thick disks.

The main purpose of this work is to investigate the possibility of
thermal equilibrium in accretion systems without outflows.
We will focus on the optically thin flows under spherical coordinates.
The paper is organized as follows.
Equations and boundary conditions are derived in Section~2.
Analyses of the vertical structure are presented in Section~3.
Numerical results of the energy advection are shown in Section~4.
The mass outflow rate is estimated in Section~5.
Conclusions and discussion are made in Section~6.

\section{Equations and boundary conditions}

Based on the assumption that there is no outflow,
we study a steady state, axisymmetric accretion flow
in spherical coordinates ($r$, $\theta$, $\phi$)
under the Newtonian potential, $\psi = -GM/r$,
where $M$ is the black hole mass.
For simplicity, we assume $v_{\theta} = 0$, which corresponds to
a hydrostatic equilibrium in the $\theta$ (or vertical) direction.
Then, the momentum equations in the $r$, $\theta$, and $\phi$
direction take the forms,
\begin{equation}
v_r \frac{d v_r}{dr} + \frac{1}{\rho}\frac{d p}{dr}
- \frac{v_{\phi}^2}{r} + \frac{GM}{r^2} = 0 \ ,
\end{equation}
\begin{equation}
\frac{1}{\rho}\frac{d p}{d\theta} - v_{\phi}^2 \cot\theta = 0 \ ,
\end{equation}
\begin{equation}
\frac{d (r v_{\phi})}{dr} - \frac{1}{r^2 \rho v_r}
\frac{d(r^3 \tau_{r\phi})}{dr} = 0 \ ,
\end{equation}
where $v_r$ and $v_{\phi}$ are respectively the radial and azimuthal
velocity, $\rho$ is the density, and $p$ is the pressure. The $r\phi$
component of the stress tensor $\tau_{r\phi} = \nu\rho r
\partial (v_{\phi}/r)/\partial r$, where $\nu = \alpha p r/\rho v_{\rm K}$
is the kinematic viscosity coefficient, and $\alpha$ is a constant
viscosity parameter.

Our main focus is the vertical structure and therefore we will
make some simplification in the radial direction.
Following \citet{NY94}, the radial self-similarity is adopted:
$v_r \propto r^{-1/2}$, $v_{\phi} \propto r^{-1/2}$,
$\rho \propto r^{-3/2}$, and $p \propto r^{-5/2}$.
We would stress that, as shown by Figures~1 and 2 in \citet{NKH97},
the self-similar solution does not perfectly match with the numerical
one, especially for the region close to the inner and outer boundaries.
Besides, the effects of general relativity are not considered in the
present work. For the innermost part of the flow, since the
gravitational force is significantly stronger in general relativity
than that in the Newtonian one, the present analysis may not apply
to this particular region.

In addition, in order to avoid detailed radiative process, we assume
a polytropic relation $p = K\rho^{\Gamma}$ in the $\theta$ direction.
We would point out a significant difference between the present work
and \citet{Gu09}. In \citet{Gu09},
$\Gamma$ was assumed to be identical to the ratio of specific heats
$\gamma$, and therefore $\Gamma$ was limited to the range [4/3,~5/3].
However, some simulations \citep[e.g., Figure~3 of][]{Villiers05}
revealed that the density may decrease faster than
the pressure from the equatorial plane to the surface, which prefers to
$\Gamma < 1$. In the present work, we will study for a wider range
[0.1,~1.7] for $\Gamma$ and take $\Gamma = 1.5$ and $\Gamma = 0.5$
as two typical examples.
With the above assumptions, the dynamic equations (1-3) are reduced to
\begin{equation}
\frac{1}{2} v_r^2 + \frac{5}{2} \lambda v_{\rm K}^2
\tilde\rho^{\Gamma-1} + v_{\phi}^2 - v_{\rm K}^2 = 0 \ ,
\end{equation}
\begin{equation}
\lambda \Gamma v_{\rm K}^2 \tilde\rho^{\Gamma-2}
\frac{d \tilde\rho}{d \theta} - v_{\phi}^2 \cot\theta = 0 \ ,
\end{equation}
\begin{equation}
v_r = - \frac{3}{2} \alpha \lambda v_{\rm K} \tilde\rho^{\Gamma-1} \ ,
\end{equation}
where $\tilde\rho \equiv \rho/\rho_0$ is the dimensionless mass density,
$\rho_0$ is the density on the equatorial plane,
$\lambda \equiv (p_0/\rho_0)/v_{\rm K}^2$, and $v_{\rm K} = (GM/r)^{1/2}$
is the Keplerian velocity. The physical meaning of $\lambda$ is the energy
advection strength on the equatorial plane, as can be indicated
by Equations~(6) and (12).

By combining Equations~(4-6) and eliminating
$v_r$ and $v_{\phi}$, we can obtain the following differential equation
for the density $\tilde\rho$,
\begin{equation}
\frac{d \tilde\rho}{d \theta} = \frac{\cot\theta}{\Gamma}
\left( \frac{\tilde\rho^{2-\Gamma}}{\lambda} - \frac{5}{2}\tilde\rho
- \frac{9}{8} \alpha^2 \lambda \tilde\rho^{\Gamma} \right) \ .
\end{equation}
Since the above first order differential equation has an unknown parameter
$\lambda$, two boundary conditions are required to solve this equation.
A natural condition is on the equatorial plane, i.e.,
$\tilde\rho|_{\theta = \pi/2} = 1$.
The other condition can be fixed at a certain polar angle $\theta_{\rm s}$,
which can be regarded as the surface of the flow.
Once the density $\tilde\rho$ at $\theta_{\rm s}$
is given, the parameter $\lambda$ as well as the vertical structure
can be derived. The numerical results will be presented in Section~4.

\section{Analyses}

Before directly solve the equations in Section~2, we
will manage to derive some analytic results.
Obviously, on the right hand side of Equation~(7), the third term
($-9 \alpha^2 \lambda \tilde\rho^{\Gamma}/8$) is significantly smaller
than the other two terms even for relatively high viscosity such as
$\alpha = 0.1$. In our analyses, this term will be dropped and Equation~(7)
is therefore reduced to
\begin{equation}
\frac{d \tilde\rho}{d \theta} = \frac{\cot\theta}{\Gamma}
\left( \frac{\tilde\rho^{2-\Gamma}}{\lambda} - \frac{5}{2}\tilde\rho
\right) \ .
\end{equation}
The above differential equation can be analytically integrated and the
vertical profile of $\tilde\rho$ is expressed by
\begin{equation}
\frac{1 - \tilde\rho^{\Gamma-1}}{1 - \tilde\rho_{\rm s}^{\Gamma-1}}
= \frac{\sin\theta^{-\frac{5(\Gamma -1)}{2\Gamma}}-1}
{\sin\theta_{\rm s}^{-\frac{5(\Gamma -1)}{2\Gamma}}-1} \ ,
\end{equation}
where $\tilde\rho_{\rm s}$ is the dimensionless density at
$\theta = \theta_{\rm s}$.
In this work we will fix $\theta_{\rm s} = \pi/6$ and
focus on the region $\pi/6 \leqslant \theta \leqslant \pi/2$.
Some simulations \citep[e.g.,][]{Villiers05,YB10}
showed $\tilde\rho \la 10^{-2}$ (or even $\la 10^{-3}$) at $\theta = \pi/6$,
which indicates that the region $\pi/6 \leqslant \theta \leqslant 5\pi/6$
will contain most of the accreted gas.
We would stress that the method here is likely to be more appropriate
than that in \citet{Gu09}, since we adopt a value for $\tilde\rho_{\rm s}$
at $\theta_{\rm s}$ according to simulation results rather than
fix $\rho = 0$ and $p = 0$ as boundary conditions no matter where they locate.

We will first study the vertical structure of the flows for $\Gamma = 1.5$
and $\Gamma = 0.5$ as two typical examples, where $\tilde\rho_{\rm s} = 0.01$
is adopted.
The vertical profiles of the density $\tilde\rho$,
the azimuthal velocity $v_{\phi}$, and the radial velocity $v_r$
are plotted in Figure~1. Both Figure~1(a) and 1(b) show that the density
decreases from the equatorial plane to the surface. The difference is that,
for $\Gamma = 1.5$, $v_{\phi}$ increases and $|v_r|$
decreases from the equatorial plane to the surface, whereas for
$\Gamma = 0.5$, the opposite behavior occurs.
Actually, the vertical profiles with $\Gamma = 1.5$ is a typical
example for the case with $\Gamma > 1$, and the profiles with $\Gamma = 0.5$
is a typical example for the case with $\Gamma < 1$.
The results can be easily understood by Equations~(4) and (6).
The latter shows $v_r \propto \tilde\rho^{\Gamma-1}$, so the profile
of $v_r$ is relevant to the sign of ($\Gamma-1$).
The former shows that both the first and second terms on the left hand side
are determined by $\tilde\rho^{\Gamma-1}$, and therefore the profile
of $v_{\phi}$ is also relevant to the sign of ($\Gamma-1$).
From another point of view, if we define a sound speed as
$c_{\rm s}^2 = p/\rho$, then there exists
$c_{\rm s}^2 \propto \rho^{\Gamma-1}$.
Since both $v_r$ and $v_{\phi}$ can be expressed by $c_{\rm s}$,
the opposite behavior of $v_r$ and $v_{\phi}$ in these two panels
is also related to the profile of $c_{\rm s}$,
thus simply the sign of ($\Gamma-1$).

The value of $\lambda$ can also be analytically derived with given
pair of ($\theta_{\rm s},\tilde\rho_{\rm s}$),
\begin{equation}
\lambda = \frac{2}{5}
\frac{\sin\theta_{\rm s}^{-\frac{5(\Gamma-1)}{2\Gamma}}-1}
{\sin\theta_{\rm s}^{-\frac{5(\Gamma-1)}{2\Gamma}}
- \tilde\rho_{\rm s}^{\Gamma -1}} \ .
\end{equation}
The variation of $\lambda$ with $\Gamma$ for $\tilde\rho_{\rm s} = 0.1$,
0.01, and 0.001 is plotted in Figure~2, which shows that $\lambda$
increases with increasing $\Gamma$ for a fixed $\tilde\rho_{\rm s}$,
and also increases with increasing $\tilde\rho_{\rm s}$ for a certain
value of $\Gamma$.

As mentioned in Section~2, the physical meaning of $\lambda$ is the
strength of energy advection on the equatorial plane.
Obviously, for a certain fixed $\tilde\rho_{\rm s}$,
Figure~2 shows that the advection on the equatorial plane increases
with increasing $\Gamma$. In other words, the advection
for $\Gamma > 1$ is stronger than that for $\Gamma < 1$ at $\theta = \pi/2$.
However, for the vertically integrated advection
from the equatorial plane to the surface, it remains uncertain which one
is stronger. The reason is that, with the relationship
$q_{\rm adv}/q_{\rm vis} \propto |v_r|/v_{\phi}^2$
(see Equations~11-12 below), since Figure~1(a) shows that $|v_r|$ decreases
and $v_{\phi}$ increases from $\theta = \pi/2$ to $\pi/6$,
the advection strength will decrease from the equatorial plane to the surface.
On the contrary, Figure~1(b) shows the opposite behavior of $|v_r|$
and $v_{\phi}$, which means that the advection strength will increase from
$\theta = \pi/2$ to $\pi/6$. Thus, the profile of the vertically
integrated advection with $\Gamma$ requires detailed numerical
calculations. We will investigate this issue in next section.

\section{Energy advection}

In this section we will focus on the strength of energy advection.
The viscous heating rate and the advective cooling rate per unit
volume take the form \citep[e.g.,][]{Gu09}:
\begin{equation}
q_{\rm vis} = \frac{9}{4}
\frac{\alpha p_0 v_{\phi}^2 \tilde\rho^{\Gamma}}{r v_{\rm K}} \ ,
\end{equation}
\begin{equation}
q_{\rm adv} = - \frac{5 - 3\gamma}{2(\gamma -1)}
\frac{p_0 v_r \tilde\rho^{\Gamma}}{r} \ ,
\end{equation}
where $\gamma$ is the ratio of specific heats. For optically thin
flows we adopt $\gamma = 1.5$, which corresponds to roughly equal amounts
of gas and magnetic pressure \citep{NY95b}.
Then, the vertical integration of the above two rates are the following:
\begin{equation}
Q_{\rm vis} = 2 \int_{\frac{\pi}{6}}^{\frac{\pi}{2}}
q_{\rm vis} \ r \sin\theta \ d\theta \ ,
\end{equation}
\begin{equation}
Q_{\rm adv} = 2 \int_{\frac{\pi}{6}}^{\frac{\pi}{2}}
q_{\rm adv} \ r \sin\theta \ d\theta \ .
\end{equation}
The energy advection factor is defined as $f_{\rm adv}
\equiv Q_{\rm adv}/Q_{\rm vis}$.
The numerical methods to obtain the variation of $f_{\rm adv}$
with $\Gamma$ are as follows. For a given value of
$\tilde\rho_{\rm s}$ at $\pi/6$ (e.g., $\tilde\rho_{\rm s} = 0.01$),
we solve Equation~(7) and derive the vertical profile of $\tilde\rho$
and the parameter $\lambda$. Then, by Equations~(4) and (6) we can
obtain the profiles of $v_r$ and $v_{\phi}$. Thus, with Equations~(11-14)
we can derive the value of $f_{\rm adv}$.
In the calculation the viscosity parameter is fixed as $\alpha = 0.1$.

The variation of $f_{\rm adv}$ with $\Gamma$
for $\tilde\rho_{\rm s} = 0.1$, 0.01, and 0.001 is plotted in Figure~3.
It is seen that the advection factor is far below unity, i.e.,
$f_{\rm adv} \la 0.3$ for $0.1 < \Gamma < 1.7$.
For the particular range $\tilde\rho_{\rm s} \la 0.01$
and $0.5 < \Gamma < 1$, which was indicated by some simulation results
\citep[e.g., Figure~3 of][]{Villiers05},
this figure shows even smaller advection factor with $f_{\rm adv} < 0.1$.
The above results illustrate that
the advective cooling itself cannot balance the viscous heating.

For optically thin flows, the radiative efficiency
$\eta~(\equiv Q_{\rm rad}/Q_{\rm vis}$, where $Q_{\rm rad}$ is the
radiative cooling rate per unit area) is roughly proportional to $\dot M$
\citep[e.g.,][chap.~9.1, p.~290]{Kato08}.
Thus, for relatively low accretion rates
(e.g., $\dot M \la 10^{-3}\dot M_{\rm Edd}$), $\eta$ will be quite small
and the radiative cooling rate will be negligible compared with
the viscous heating rate. Consequently, the viscous heating rate
will be larger than the sum of the advective cooling rate and the radiative
cooling one. In other words, no thermal equilibrium can be established
under this situation. The contradiction may come from the original
assumption that no outflow exists in the accretion system. We therefore
propose that outflows ought to occur in optically thin flows with
low accretion rates such as
$\dot M \la 10^{-3}\dot M_{\rm Edd}$. Moreover, the outflows should
take away more than 70\% of the thermal energy generated by viscous
dissipation.
As mentioned in Section~1, outflows have been found in accretion system
by both observations and numerical simulations.
Our analyses and numerical results may help to understand the mechanism,
which is probably related to the insufficient energy advection.

\section{Mass outflow rate}

In this section we will estimate the total mass outflow rate
based on the advection factor $f_{\rm adv}$. We define the
quantity $\dot{M}_{\rm in}$ as the mass accretion rate of the inflow,
which will increase with increasing $r$.
For a small range $(r_0,~r_0+\Delta r)$,
the mass outflow rate from this specific region can be expressed as
\begin{equation}
\Delta \dot{M}_{\rm out} = \dot{M}_{\rm in}|_{r=r_0+\Delta r}
- \dot{M}_{\rm in}|_{r=r_0} \ .
\end{equation}
Obviously, the total mass outflow rate can be written as
\begin{equation}
\dot{M}_{\rm out} = \dot{M}_{\rm in}|_{r=r_{\rm out}}
- \dot{M}_{\rm in}|_{r=r_{\rm in}} \ ,
\end{equation}
where $r_{\rm in}$ and $r_{\rm out}$ are the inner and outer
boundary, respectively.
The above two equations show that, once the radial profile of
$\dot{M}_{\rm in}$ is determined, the profile of mass outflow rate
together with the total rate $\dot{M}_{\rm out}$ can be derived.

According to the spirit of this work, if the advective cooling could balance
the viscous heating, i.e., $f_{\rm adv} = 1$, then the thermal equilibrium
could be established and therefore it is not necessary for the occurrence
of outflow.
In other words, $\dot{M}_{\rm in}$ could be a constant if $f_{\rm adv} = 1$
were realized.
We therefore assume that the inflow rate $\dot {M}_{\rm in}$
will satisfy the following equation:
\begin{equation}
\frac{d\ln \dot {M}_{\rm in}}{d\ln r} = \delta (1-f_{\rm adv}) \ ,
\end{equation}
where $\delta \la 1$ is a dimensionless parameter. By the radial integration
of the above equation we can obtain the ratio of inflow rate
at the inner boundary to that at the outer boundary,
\begin{equation}
\frac{\dot{M}_{\rm in}|_{r=r_{\rm in}}}{\dot{M}_{\rm in}|_{r=r_{\rm out}}}
= \left( \frac{r_{\rm in}}{r_{\rm out}} \right)^{\delta (1-f_{\rm adv})}
\ .
\end{equation}

The numerical results are shown in Figure~4,
where $r_{\rm in} = 3 r_{\rm g}$ and $r_{\rm out} = 10^3 r_{\rm g}$.
The variation of such a ratio with $\Gamma$ is plotted for
$\delta = 1$, $1/2$, and $1/3$.
It is seen that the ratio is not sensitive to $\Gamma$, but decreases
rapidly with increasing $\delta$, as can be inferred by
Equation~(18). The low values of this ratio implies that the outflow
will be significant and even dominant, which agrees with the observation
that only a small part of the original gas can be accreted by the
supermassive black hole in the Milky Way.
For a comparison with the numerical simulations of \citet{Yuan12},
as mentioned in Section~1, the inflow index $s$ is in the range
$[0.4,~0.7]$, which indicates that $\delta$ is probably in the range
$[0.4,~1]$ (according to Equation~17) since our results show
$f_{\rm adv} \la 0.3$.
In our understanding, the physical reason for $\delta$ less than unity
may be related to the radiative cooling, which can balance a part of
the viscous heating particularly for high accretion rates.

\section{Conclusions and discussion}

In the present paper, we have investigated the steady state, axisymmetric,
optically thin accretion flows in spherical coordinates
under the no-outflow assumption.
We have found that the advective cooling rate is generally less than
30\% of the viscous heating rate.
As a consequence, for radiatively inefficient flows with low accretion
rates such as $\dot M \la 10^{-3}\dot M_{\rm Edd}$, no thermal equilibrium
can be established since the viscous heating rate will be larger
than the sum of the advective cooling rate and the radiative cooling one.
We therefore argue that in such case outflows ought to occur and take away
more than 70\% of the thermal energy generated by viscous dissipation.
Our results may help to understand the mechanism of outflows found in
observations and numerical simulations.

On the other hand, for optically thick flows with extremely high accretion
rates (e.g., $\dot M \ga 10\dot M_{\rm Edd}$), the viscous heating is
so strong that the radiative cooling will be quite small compared with
the heating. The classic slim disk model predicts that the advective cooling
will balance the viscous heating once the half-thickness of the disk $H$
approaches the cylindrical radius $R$. However, \citet{Gu12} studied
radiation pressure-supported disks with radiative transfer and showed that
the disk will be extremely thick when advection is dominant.
Moreover, for extremely high accretion rates, no thermal equilibrium
solution was found, which also implies the occurrence of outflows.
Here, it is easy to estimate the advection strength for optical thick flows
by Equation~(12) and Figure~3. The main difference between optically thin
and thick cases is related to $\gamma$, which is likely to be 4/3
for radiation pressure-dominated flows.
In Section~4 we adopt $\gamma = 3/2$ for optically thin flows.
Equation~(12) implies that $q_{\rm adv}$ will be three times larger for
$\gamma = 4/3$ than that for $\gamma = 3/2$. Thus, we can expect
$f_{\rm adv} \la 0.9$ for $0.1 < \Gamma < 1.7$. Moreover, with Figure~3
we can even expect $f_{\rm adv} < 0.2$ for the particular range
$\tilde\rho_{\rm s} \la 0.01$ and $0.5 < \Gamma < 1$,
which is preferred according to simulation results. Consequently,
under the no-outflow assumption, there will be no thermal equilibrium
either for $\dot M \ga 10\dot M_{\rm Edd}$.
In other words, in such case outflows should also occur and take away
the excess thermal energy, which is generated by viscous dissipation
and cannot be balanced by advection plus radiation.

In our opinion, the strength of energy advection is a key point in
accretion theory. Most previous works on this issue can
be classified as the following two types. The first one is under cylindrical
coordinates by using a Taylor expansion of gravitational force in the
vertical direction, or equivalently by using a simple relation,
i.e., $H = c_{\rm s}/\Omega_{\rm K}$,
where $\Omega_{\rm K}$ is the Keplerian angular velocity
\citep[e.g.,][]{Abram88,Wang99,Watar00,Sadow11}.
Such an approach may significantly magnify the original force and therefore
greatly enlarge the strength of energy advection. The second one is under
spherical coordinates by fixing a value of the local advection factor
($f'_{\rm adv} \equiv q_{\rm adv}/q_{\rm vis}$) in advance
\citep[e.g.,][]{NY95a,Xu97,Xue05,Jiao11}. We would stress
that both of the above two approaches may not properly derive
the real strength of the advection. The method in the present work,
however, provides a possible clue to investigate such a strength.

The present work is based on the constant $\alpha$ assumption.
As shown by some simulations \citep[e.g.,][]{Hirose09,Jiang14},
however, the parameter $\alpha$ is not likely to be a constant
in the vertical direction, which is probably relevant to
the magnetic pressure. For instance, there may exist magnetically
dominated coronae above the flow and therefore the constant
$\alpha$ assumption may be violated. In addition, convection
is not considered in this work, which may transfer energy
in the vertical or radial direction. Thus, the above two issues
may have quantitative influence on the present results.

As mentioned in the first section, outflows may be driven by several
mechanisms such as radiation pressure and magnetic fields. In this work,
we avoid the detailed mechanism to power outflows. Our no thermal equilibrium
existence can be regarded as a necessary condition to illustrate
that outflows are inevitable for both optically thin flows with low accretion
rates and optically thick flows with extremely high accretion rates.
In our opinion, it is worthy to check the strength of energy advection
in simulations, particularly for those with strong outflows.
Such a work may not be difficult since the values of physical quantities
in Equations~(11-14) can be derived through simulations.

\acknowledgments

The author thanks Feng Yuan and De-Fu Bu for beneficial discussion,
and the referee for helpful comments to improve the paper.
This work was supported by the National Basic Research Program of China
(973 Program) under grant 2014CB845800, the National Natural Science
Foundation of China under grants 11222328 and 11333004,
and the Fundamental Research Funds for the Central Universities
under grant 20720140532.

\clearpage

\begin{figure}
\plottwo{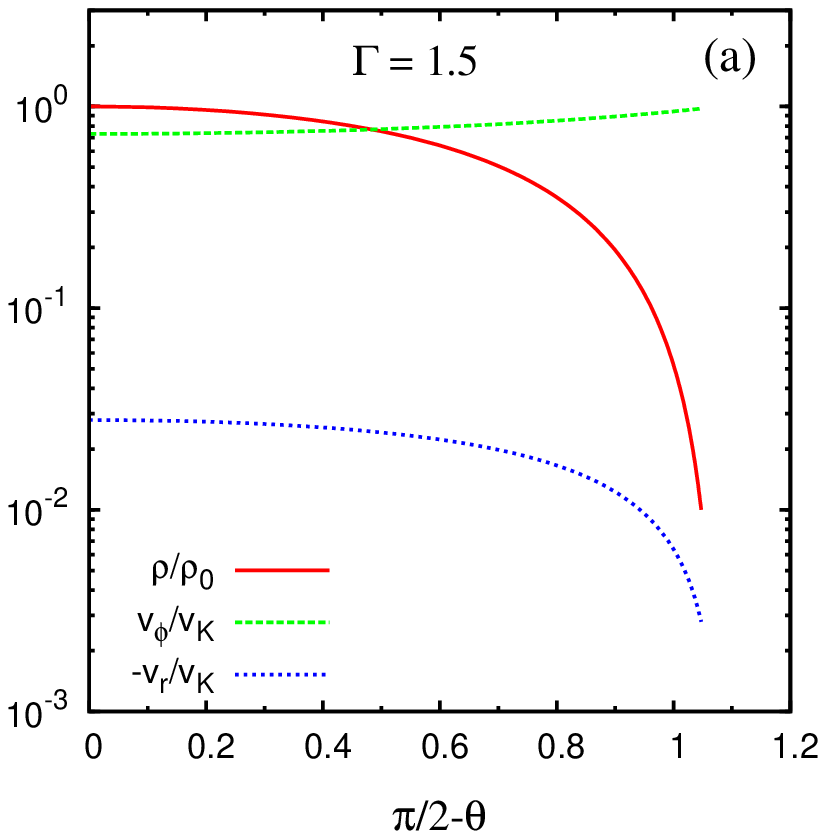}{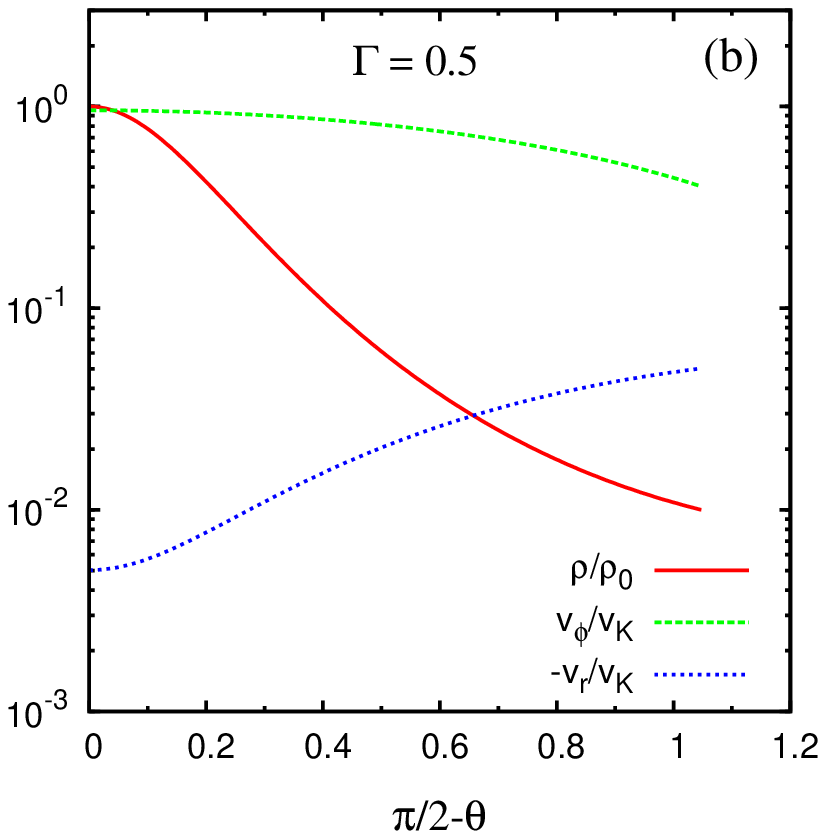}
\caption{
Variations of $\rho$, $v_{\phi}$, and $v_r$ for $\tilde\rho_{\rm s} = 0.01$:
(a) $\Gamma = 1.5$; (b) $\Gamma = 0.5$.
}
\end{figure}

\clearpage

\begin{figure}
\plotone{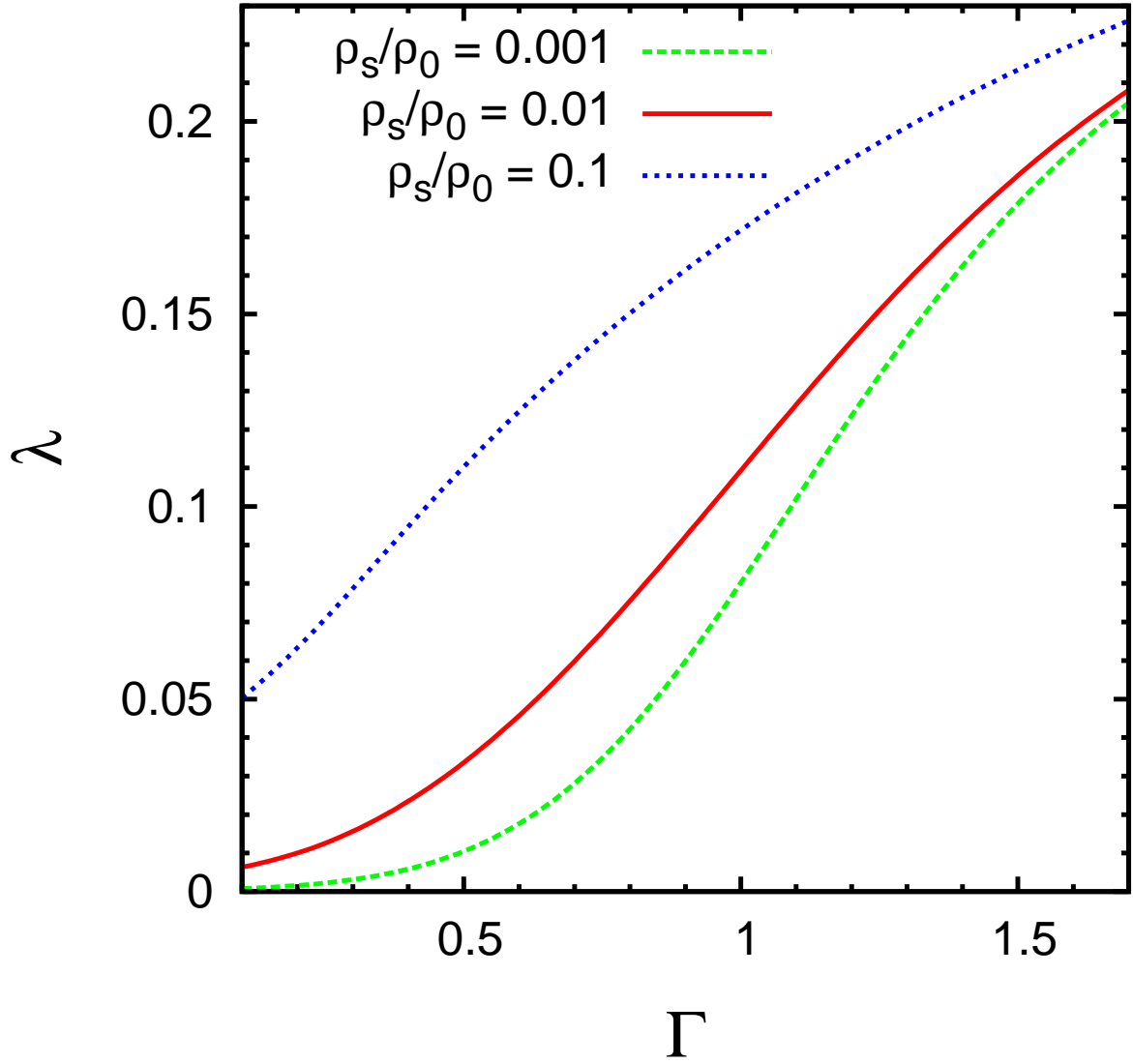}
\caption{
Variation of $\lambda$ with $\Gamma$ for $\tilde\rho_{\rm s} = 0.1$,
0.01, and 0.001.
}
\end{figure}

\clearpage

\begin{figure}
\plotone{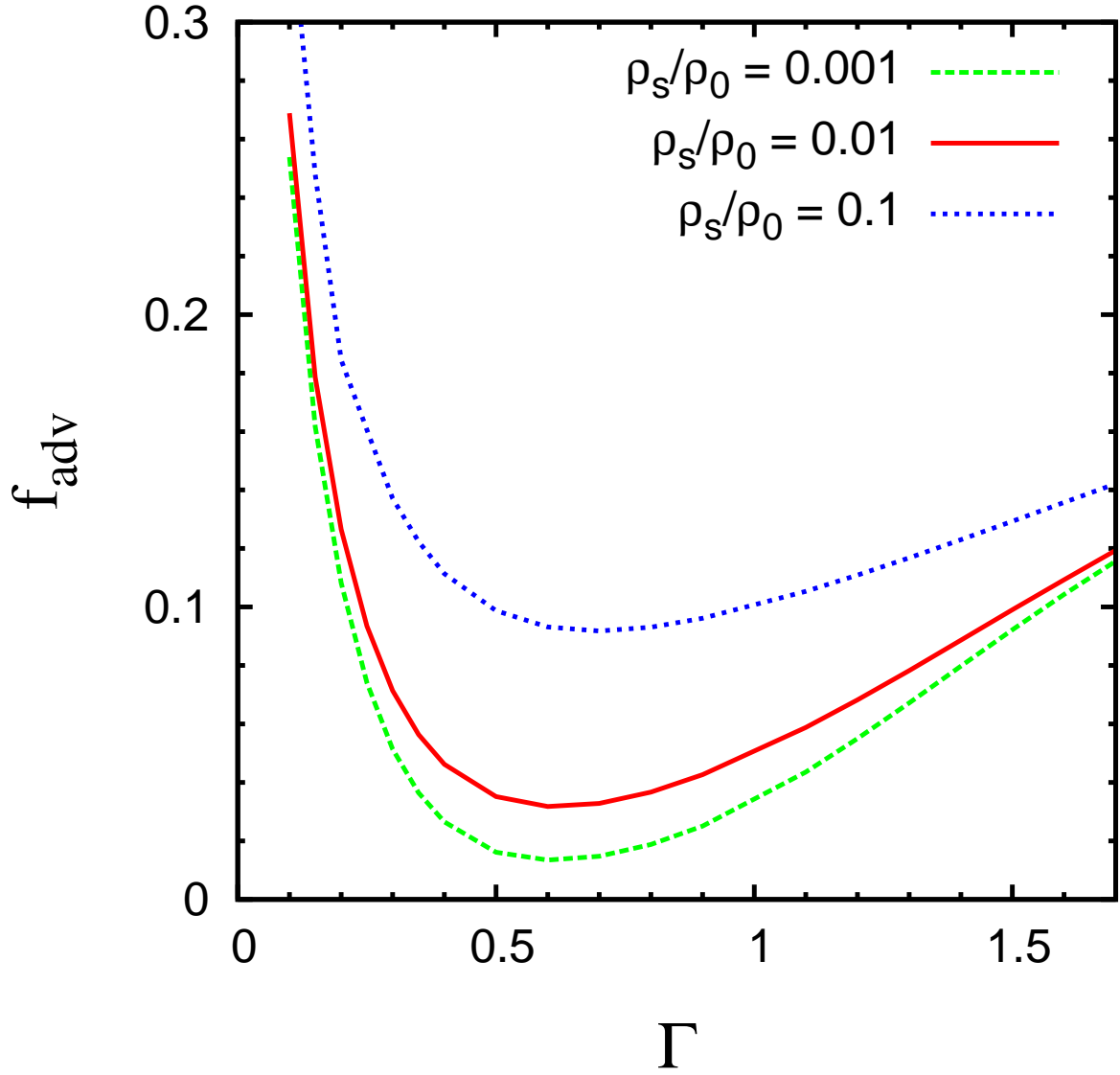}
\caption{
Variation of the advection factor $f_{\rm adv}$ with $\Gamma$ for
$\tilde\rho_{\rm s} = 0.1$, 0.01, and 0.001.
}
\end{figure}

\clearpage

\begin{figure}
\plotone{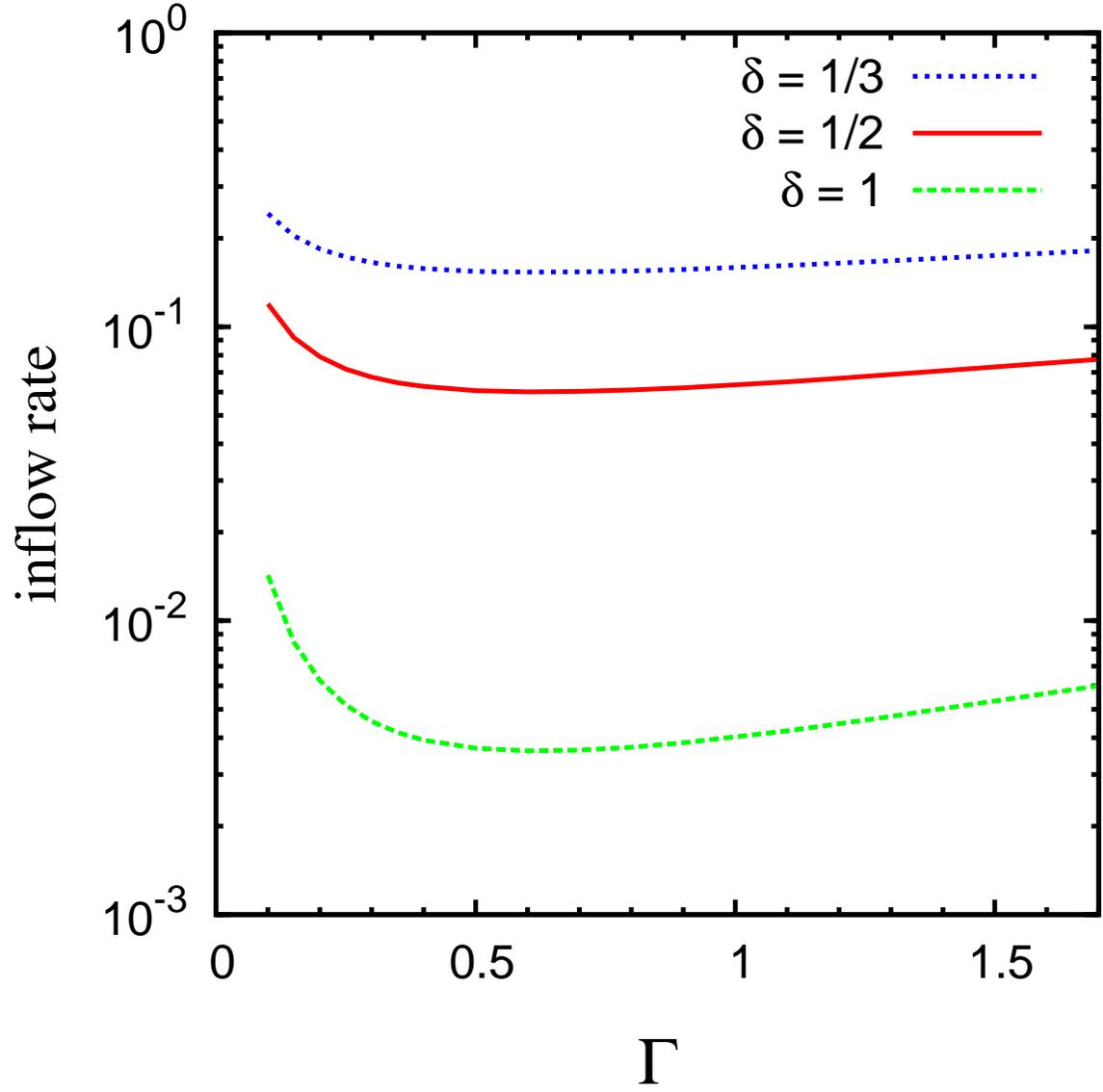}
\caption{
Variation of the ratio of inflow rate
$(\dot{M}_{\rm in}|_{r=3 r_{\rm g}})/(\dot{M}_{\rm in}|_{r=10^3 r_{\rm g}})$
with $\Gamma$ for $\delta = 1$, 1/2, and 1/3.
}
\end{figure}

\end{document}